\documentclass[%
 aip, amsmath,amssymb,reprint]{revtex4-1}

\newcommand{\ra}{\rangle}

\newcommand{\upa}{\uparrow}
\newcommand{\dna}{\downarrow}
\usepackage{color}
\usepackage{graphicx}
\usepackage{epsfig}
\usepackage{ulem}

\begin{document}

\title{Device/circuit simulations of silicon spin qubits based on a gate-all-around transistor}

\author{Tetsufumi Tanamoto}
\affiliation{Department of Information and Electronic Engineering, Teikyo University,
Toyosatodai, Utsunomiya  320-8511, Japan} 
\email{tanamoto@ics.teikyo-u.ac.jp}

\author{Keiji Ono}
\affiliation{Advanced device laboratory, RIKEN, Wako-shi, Saitama 351-0198, Japan}

\begin{abstract}
We theoretically investigated the readout process of a spin--qubit structure based on a gate-all-around (GAA) transistor. 
Our study focuses on a logical qubit composed of two physical qubits. 
Different spin configurations result in different charge distributions, 
which subsequently influence the electrostatic effects on the GAA transistor. 
Consequently, the current flowing through the GAA transistor depends on the qubit's state.
We calculated the current-voltage characteristics 
of the three-dimensional configurations of the qubit and GAA structures, 
using technology computer-aided design (TCAD) simulations.
Moreover, we performed circuit simulations using the Simulation Program with Integrated Circuit Emphasis (SPICE) 
to investigate whether a readout circuit made from complementary metal--oxide semiconductor (CMOS) transistors can amplify the weak signals generated by the qubits. 
Our findings indicate that, by dynamically controlling the applied voltage within a properly designed circuit, 
the readout can be detected effectively based on a conventional sense amplifier.
\end{abstract}

\maketitle

\section{Introduction}
\label{sec:introduction}
The development of quantum computing devices is one of the most active area of contemporary research. 
Semiconductor qubits have the advantage of being implemented directly into conventional 
complementary metal--oxide semiconductor (CMOS) circuits~\cite{Stuyck,Anders}. 
Because of their small size, they are promising for large-scale integration. 
Currently, the most advanced commercial transistors feature a gate length of approximately 2 nm
which has been announced for use in smart-phones and personal computers~\cite{Yeap}.
In addition, CMOS circuits are being constructed as three-dimensional (3D) structures~\cite{Loubet,Bae,Agrawal,Jung,Sreenivasulu,Chen,Pal,Abdi,Lee}. 

The sizes of quantum dots, which are elements of semiconductor qubits, 
are on the order of approximately 50 nm~\cite{Zajac,Yoneda,TaruchaCNOT}.
These sizes are significantly larger than the gate lengths of recently developed commercial transistors. 
Moreover, advanced transistors, such as those used in NAND flash memories, exhibit high sensitivity 
to local charge variations, owing to their reduced dimensions and optimized channel structures~\cite{Monzio}. 
This suggests that small transistors fabricated via industrial processes may be 
capable of detecting spin-dependent charge distributions of qubits, as explained below.

{\it Spin Qubit}---
The spin degrees of freedom ($\uparrow$ and $\downarrow$-spins) have a sufficiently long coherence time
for the detection of spin states using semiconductor devices.
Thus, qubits that use charge spins are promising for quantum computers.
Loss and DiVincenzo proposed the basic concept of a spin qubit 
where an electron in a quantum dot is assigned as a qubit~\cite{Loss,Burkard}. 
This concept has been extended in several directions to create {\it logical qubits} comprising two or three quantum dots. 
One advantage of a logical qubit is its ability to convert spin degrees of freedom into charge degrees of freedom during the measurement~\cite{Taylor1,Taylor2}.
The singlet-triplet state is a typical logical qubit that has been investigated by many researchers~\cite{Petta,Taylor1,Taylor2,Maune,Klinovaja,Calderon,Takeda,Zhang}.
Taylor \textit{et al.}~\cite{Taylor1,Taylor2} conducted extensive research on singlet-triplet states. 
When an appropriate voltage was applied across the two quantum dots, 
two electrons occupy a single quantum dot with a lower energy level only if they have opposite spin directions, in accordance with the Pauli exclusion principle, 
as shown in Fig.1. 
When two electrons have the same spin direction (Fig. 1(a)), the electron in the lower quantum dot cannot move because of the Pauli exclusion principle. 
Conversely, when the two electrons have different spin directions (Fig. 1(b)), 
the electron in the lower quantum dot can be transferred to the upper quantum dot. 
Consequently, the electron distribution depends on the spin configuration after voltage is applied.

\begin{figure}
\centering
\includegraphics[width=8cm]{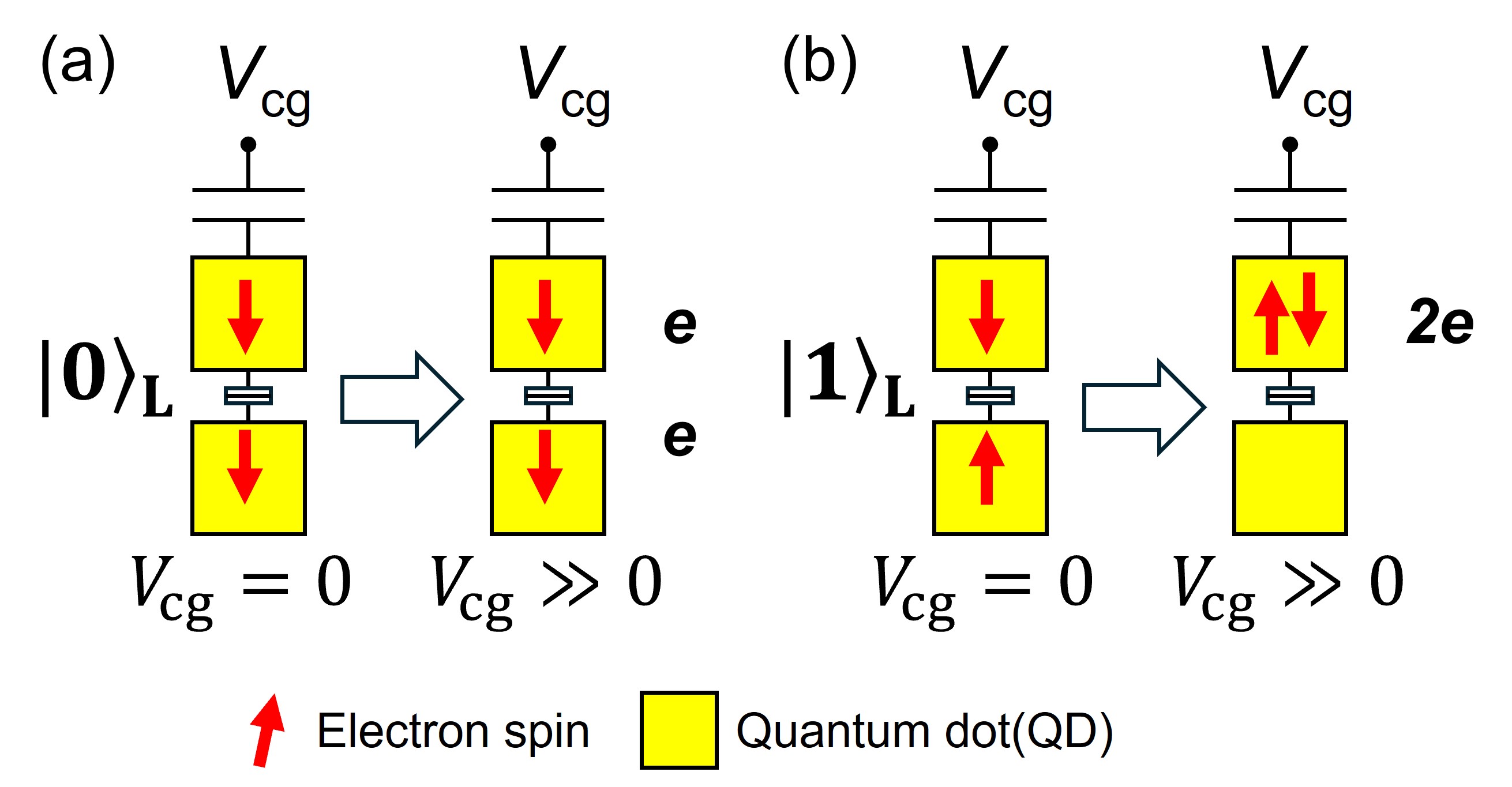}
\caption{A logical qubit is defined using two coupled quantum dots that incorporate electric spins. 
In the readout mode, as shown in the figures on the right of (a) and (b), 
a gate bias \( V_{\rm cg} \) is applied to the  quantum dots (here, upper quantum dot). 
The electrons accumulate in the upper quantum dot only 
when the logical qubit is in the \( |1\rangle_L \) state owing to the Pauli exclusion principle.
}
\label{fig1}
\end{figure}
{\it Charge Sensor}---
The spin states in double quantum dots, converted into charge distributions, 
enable electrical readout using highly sensitive charge sensors\cite{Ono,Yoneda}. 
Typical charge sensors consist of a quantum dot with their own source and drain, 
operating through capacitive coupling with nearby qubits~\cite{Burkard}.
Generally, qubit control and readout are performed using different electrodes~\cite{Yoneda,Zajac,Takeda,Maune,TaruchaCNOT}.
Therefore, the qubits are spaced apart on the substrate, 
and many wires around a single qubit are required.
The interaction strength between qubits, based on Heisenberg coupling, 
rapidly decreases as they separate.
Thus, it is essential for the spin qubits to be positioned closely together. 
In addition, to maintain the coherence of the qubits, 
close placement of qubits, e.g., a two-dimensional (2D) array, is desirable.
Hence, the general configuration of spin qubits prevents effective integration of the qubit system. 

{\it Compact Spin Qubit System}---
In \cite{TanaJAP2023,TanaJAP2025}, we proposed a compact qubit structure, 
where the quantum dots designated as qubits are placed next to the channel quantum dots. 
The channel quantum dots serve a dual purpose: they mediate the interactions between qubits and facilitate the measurement of qubit states.
The small area of the compact qubits also aligns with the economic need to miniaturize CMOS circuits.
However, the detectors described in \cite{TanaJAP2023,TanaJAP2025} consist of quantum dots,
which are different structures from the conventional transistors, 
and their fabrication requires the development of new processes.
Generally, the fabrication of small transistors is expensive, 
making it difficult for most research institutes 
and universities to access the latest semiconductor technologies. 
One possible approach is to use the commercial transistor fabrication process 
as much as possible to construct the qubit system.

{\it Readout by GAA transistor}---
In this study, we theoretically investigated a scenario in which charge sensors are replaced by gate-all-around (GAA) transistors. 
Utilizing the inherent charge sensitivity of advanced transistors, 
we aim to simplify the qubit readout architecture. 
We assessed whether GAA transistors can replace charge sensors and offer a promising pathway for CMOS-compatible spin--qubit systems.
This approach not only reduces the number of specialized components and enhances cost efficiency and reliability, but also facilitates the construction of dense 2D qubit arrays. 

The contributions of this study are summarized as follows:
\begin{itemize}
\item A spin-qubit system based on GAA transistors is proposed to construct qubits alongside advanced transistors.
\item Technology computer-aided design (TCAD) simulations demonstrate that the current-voltage characteristics of GAA transistor reflect the charge distributions of the qubit states, 
confirming the effectiveness of the GAA transistor as a readout apparatus.
\item Circuit simulations indicate that a spin qubit based on GAA transistors can be read out by modifying standard sense amplifier circuits, 
presenting a potential integration of qubit systems into conventional CMOS technology.
\end{itemize}

The remainder of this paper is organized as follows.
In Section~\ref{sec:method}, we explain our method. 
Section~\ref{sec:model} presents our basic model. 
Section~\ref{sec:TCAD} presents the numerical results of the TCAD simulations.
Section~\ref{sec:SPICE} presents the results of the circuit simulations.
Section~\ref{sec:discussions} provides discussions regarding our findings.
Finally, Section~\ref{sec:conclusions} summarizes and concludes the paper.
The appendix provides additional explanations to complement the main content.

\section{Method}\label{sec:method}
{\it Hardware Simulations}---
The use of GAA transistors raises an important question: can these transistors be effectively controlled by the gate electrode in the presence of qubits? 
Moreover, is it possible to accurately readout the qubit states through the channel current?
To answer these questions, we performed 3D calculations based on the geometric device structure 
of GAA transistor.
Generally, hardware-aware simulations of quantum computers involve more elements than conventional CMOS circuits. 
Quantum--computing research should encompass a wide range of simulations, 
including those at the physical~\cite{Johansson}, device~\cite{Ding,Birner,Beaudoin,Shehata}, and circuit levels~\cite{Rijs,Gys,TanaAPL}. 
Although several attempts have been made to develop a unified software toolchain~\cite{Mohiyaddin,Costa},
the appropriate software for designing qubit systems must be selected according to individual requirements. 
We opted for conventional TCAD and 
Simulation Program with Integrated Circuit Emphasis (SPICE) simulators, 
because these tools are readily accessible to most semiconductor engineers in their daily work.

{\it TCAD Simulations}---
TCAD is a simulation framework that numerically solves semiconductor physics equations (e.g. Poisson and drift-diffusion) on realistic device structures to predict transistor behavior before fabrication.
In this study, we employed the Silvaco TCAD simulator to analyze how excess charge affects the channel current and how significantly the channel current changes because of nearby quantum dots~\cite{TanaTCAD}. 
We must determine whether the GAA transistor can detect changes in the charge distribution of the nearby quantum dots. 
For the sake of simplicity and due to software limitations, we represented both the quantum dots and the channel region 
as square structures. 
The quantum dots were modeled as particles of silicon nitride (SiN) with uniform charge distribution. 
Although gate insulators usually include SiO${}_2$ and Hf oxides, 
the gate insulators in the present calculations are assumed to consist only of SiO${}_2$ for simplicity.
We also neglected quantum tunneling and focused on the effects of excess charge on the channel current.

{\it SPICE Simulations}---
We also explored the possibility of detecting current from the qubit array using CMOS circuits. 
Here, we employed the conventional circuit simulator SPICE (Silvaco SmartSpice) along with Verilog-A to analyze the numerical results obtained from TCAD simulations. 
Verilog-A is a high-level modeling language for analog circuits designed to describe device behavior using equations and procedural constructs.
Verilog-A can also include the dataset in the SPICE simulation.
We demonstrated that it is feasible to detect differences between qubit states using conventional CMOS circuits.

\begin{figure}
\centering
\includegraphics[width=8cm]{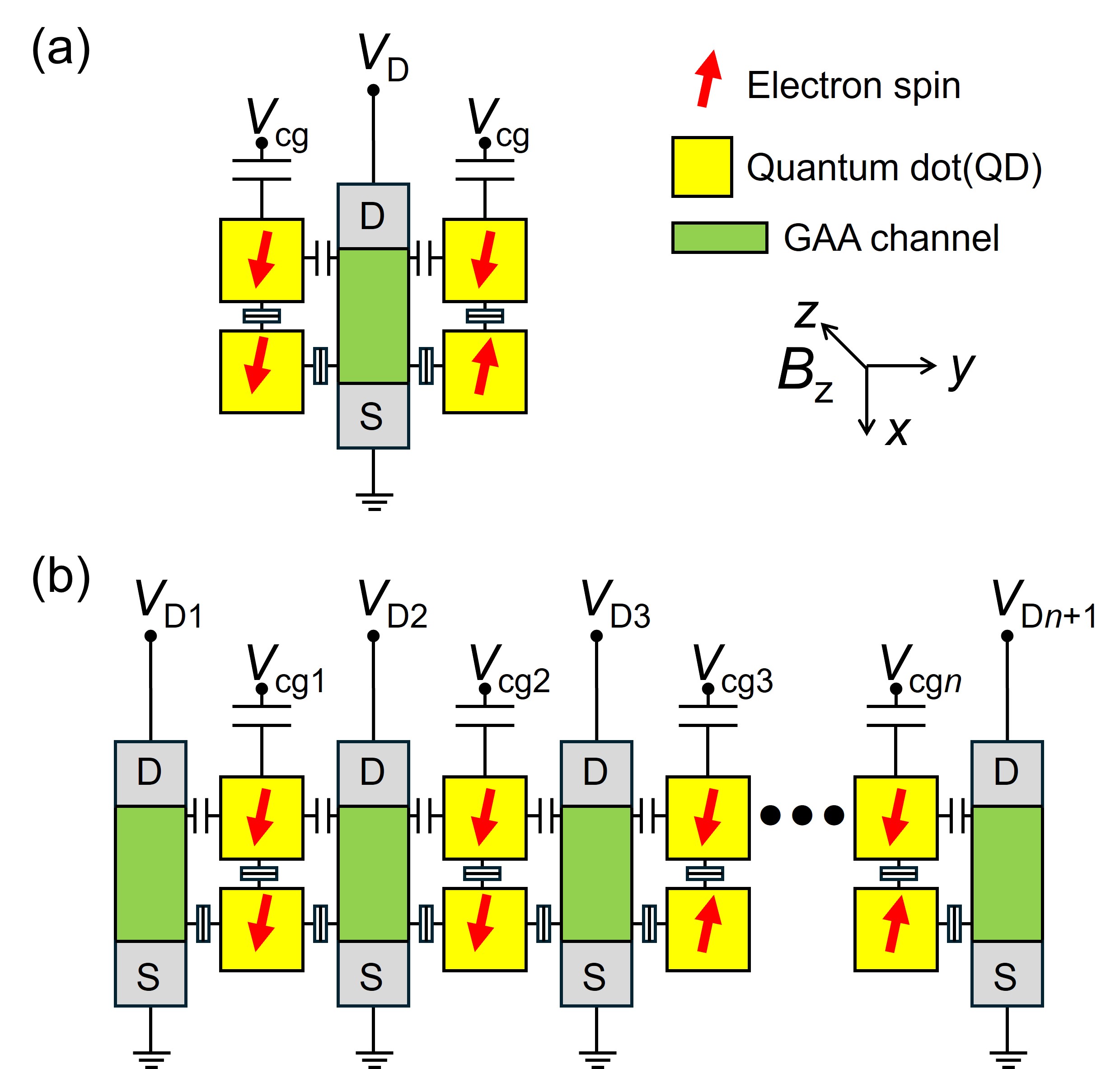}
\caption{(a) A unit of the qubit (two quantum dots) and the GAA transistor.
The qubits made of two quantum dots (Fig. 1) are interconnected through GAA transistors. 
One of the double quantum dots is tunnel-coupled to the transistor channel, 
while the other is capacitively coupled to the GAA transistors.
(b) A qubit-transistor array with the structure shown in Fig.(a). }
\label{fig2}
\end{figure}
\begin{figure}
\centering
\includegraphics[width=8.8cm]{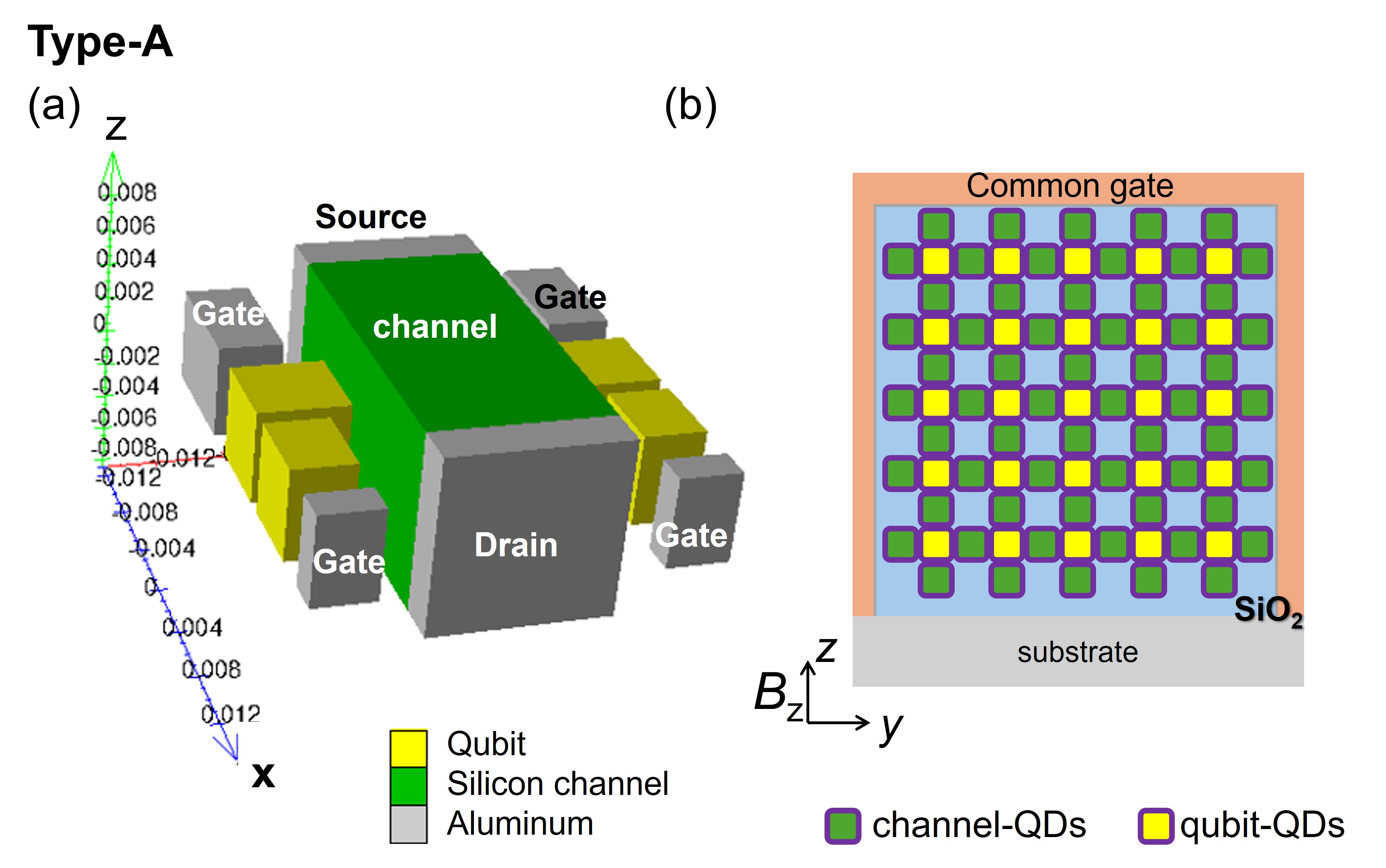}
\caption{
An example of stacking the qubit arrays of Fig.~\ref{fig2}. 
We call this structure "type-A". 
(a) A single unit. (b) 2D qubit array of (a). 
In this case, the GAA structure (green) is surrounded by two qubits (yellow) on two sides ((a)).
In the appendix \ref{sec:Appendix_A}, a different configuration (type-B) is discussed.
The qubits and GAA transistor are surrounded by the common gate via SiO${}_2$, 
which is not shown in the figure.  
}
\label{fig3}
\end{figure}

\section{Model}\label{sec:model}
\subsection{Logical qubits and GAA transistor}
Figure 2(a) shows a single unit of the proposed qubits with the GAA structure. 
Figure 2(b) shows a one-dimensional qubit array.
Quantum dots can be made by floating gates in legacy flash memory, as described in \cite{Monzio}. 
A static magnetic field $B_z$ is applied to generate Zeeman splitting.
We assume that the channel regions of the GAA transistors are surrounded by qubits 
with the same tunneling mechanism described in \cite{TanaAIP,TanaJAP2023,TanaJAP2025}. 
In Fig.2(b), the end GAA transistors monitor the nearest one-qubit states.
The GAA transistors located between the end GAAs monitor the quantum states of adjacent two qubits. 
By detecting the currents across all GAA transistors, we can infer the qubit states.
The advantage of using a field-effect transistor is its ability to detect changes in the distributions 
by using an electric potential. 

Logical qubits can be constructed, as shown in Fig.1.
The logical qubit states are defined as 
$|0\rangle_L \equiv |\downarrow\downarrow\rangle$ and 
$|1\rangle_L \equiv |\uparrow\downarrow\rangle$, as in \cite{TaruchaCNOT}.
To gain a more detailed understanding of the logical states, 
we solved the eigenvalue problem for the four coupled quantum dots. 
Exact logical states are presented in Appendix~\ref{sec:Appendix_B}.
The electric potential of the qubits is controlled by the control gate $V_{\rm cg}$.
The electric potential of the GAA transistor can also be changed by adjusting the source and drain voltage $V_{\rm D}$.

Qubit operations are performed in the same way as those in \cite{TanaJAP2023,TanaJAP2025},
where two modes are available: coupling ($V_{\rm D}=0$) and readout ($V_{\rm D}\neq 0$). 
In Fig.2, the GAA plays the role of coupling between the qubits and the readout.
The electron--electron interactions occur between the lower quantum dots,
through the electrons in the channel~\cite{Ruderman}\cite{Kasuya}\cite{Yosida}. 
An analysis regarding to the magnitude of this interaction is presented in Appendix~\ref{sec:Appendix_C}.
In the following, we focus on the readout process without treating the qubit operations.

\begin{figure}
\centering
\includegraphics[width=8.8cm]{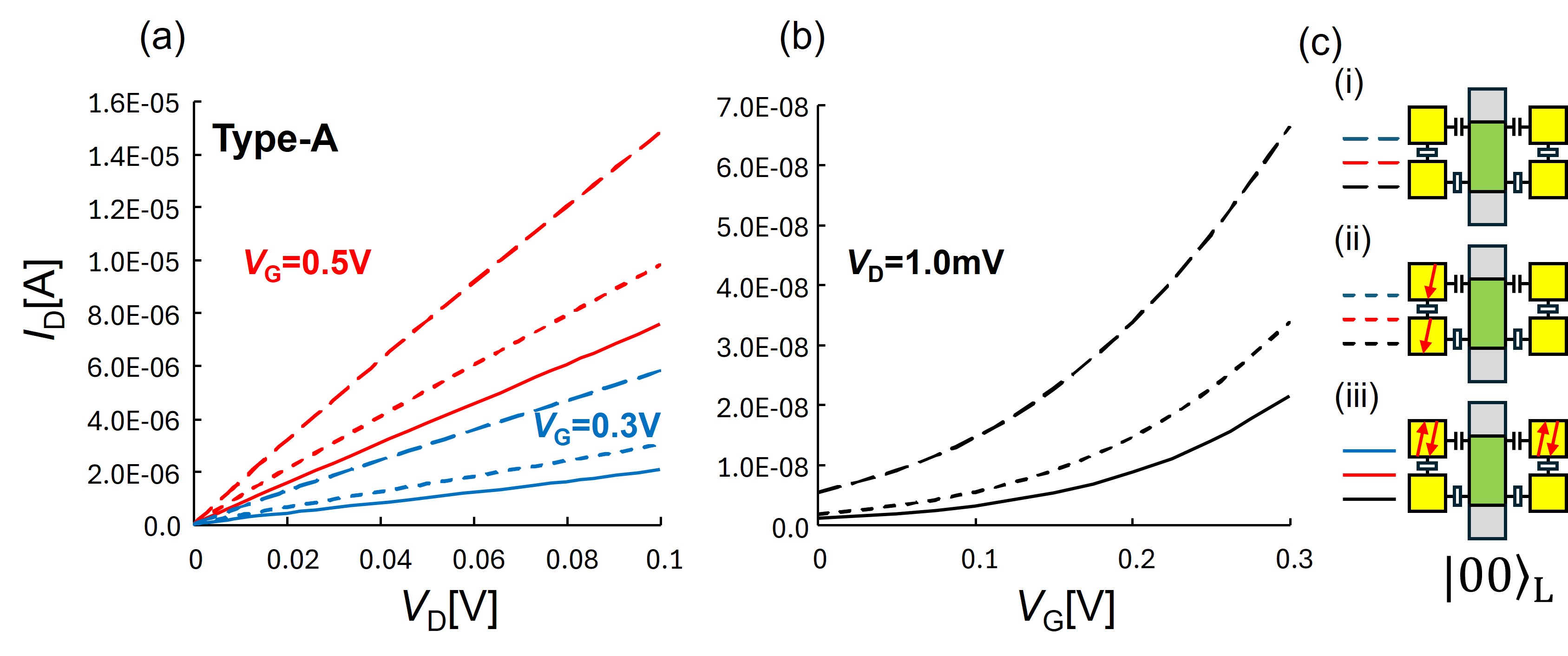}
\caption{Current-voltage characteristics of the type-A configuration
in the initialization step to confirm whether two electrons exist in 
two coupled quantum dots. 
(a) $I_D$-$V_D$ and (b) $I_D$-$V_G$ characteristics
for the 2.5~nm quantum dot qubit.
(c) Three charge distributions in (a) and (b).}
\label{fig_plane_pre}
\end{figure}

\begin{figure}
\centering
\includegraphics[width=8.8cm]{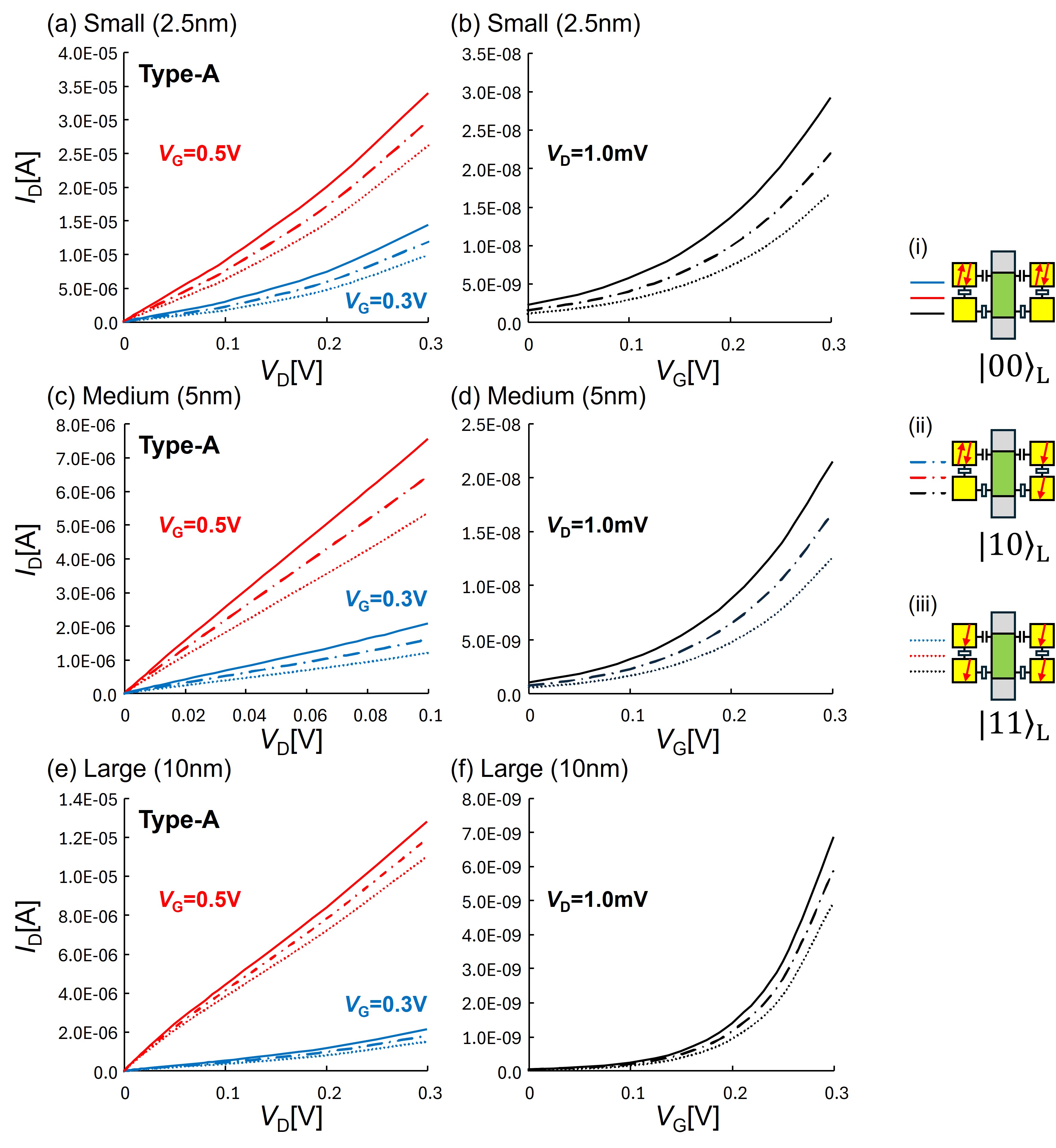}
\caption{
Current-voltage characteristics of the type-A configuration.
See Table 1 for the parameters.
(a) $I_D$-$V_D$ and (b) $I_D$-$V_G$ characteristics
for the 2.5 nm quantum dot qubit.
(c) $I_D$-$V_D$ and (d) $I_D$-$V_G$ characteristics
for the 5 nm quantum dot qubit.
(e) $I_D$-$V_D$ and (f) $I_D$-$V_G$ characteristics
for the 10 nm quantum dot qubit.
}
\label{fig_plane_distributions}
\end{figure}

\begin{table}
\caption{\textbf{TCAD parameters; "standard," "large," and "small" correspond 
to the numerical results of Fig.\ref{fig_plane_distributions}.}}
\label{table1}
\begin{center}
\begin{tabular}{|p{63pt}|p{43pt}|p{46pt}|p{45pt}|}
\hline
 &small(2.5nm) & medium(5nm) & large(10nm)   \\
\hline
Upper quantum dot & 2.5$\times$1.5$\times$2.5 nm${}^3$ & 5$\times$3$\times$5 nm${}^3$ & 10$\times$8.5$\times$10 nm${}^3$  \\
\hline
Lower quantum dot & 2.5$\times$2.5$\times$2.5 nm${}^3$  & 5$\times$5$\times$5 nm${}^3$ & 10$\times$10$\times$10 nm${}^3$ \\
\hline
Gate length & 17nm &19nm &22nm    \\
\hline
Gate width  &\multicolumn{3}{c|}{ 10$\times$10 nm${}^2$  } \\
\hline
Gate work function  &\multicolumn{3}{c|}{4.65 eV}\\
\hline
Qubit permittivity & \multicolumn{3}{c|}{11.8} \\
\hline
P-type doping & \multicolumn{3}{c|}{$10^{14}$ /cm${}^3$} \\
\hline
N-type doping & \multicolumn{3}{c|}{$10^{19}$ /cm${}^3$} \\
\hline
\end{tabular}
\end{center}
\end{table}

\subsection{2D qubit array}
When a 2D qubit array of Fig.~2 is considered, 
two different stacking structures appear, which we call the type-A and type-B configurations.
Figure~\ref{fig3}(a) shows the type-A configuration,  
in which the GAA channel is surrounded by two logical qubits.
This results in Fig.~\ref{fig3}(b),
where each qubit is surrounded by four GAA transistors. 
In this configuration, the readout current reflects the charge distribution of the two qubits.
The quantum dots close to the drain are assumed to be smaller than the lower quantum dots. 
This assumption is based on the idea that tunneling between the upper quantum dots and the GAA transistor is prohibited. 
The type-B configuration 
where the GAA channel is surrounded by four logical qubits 
is discussed in Appendix \ref{sec:Appendix_A}. 
In the type-B configuration, each qubit is surrounded by two GAA transistors.
A common gate is assumed to cover both the qubits and GAA channels. 
These designs are forward-compatible with 2D quantum error-correcting codes~\cite{Fowler}.
Hereinafter, we focus on type-A configuration.

Parameters are presented in Table 1. 
The calculations were conducted using the Silvaco Atlas simulator at room temperature. 
This was because the low-temperature calculation failed to converge owing to the complex structure. 
Additionally, for simplicity, it was assumed that there is no tunneling between the quantum dots and the channel.

\begin{figure}
\centering
\includegraphics[width=8.0cm]{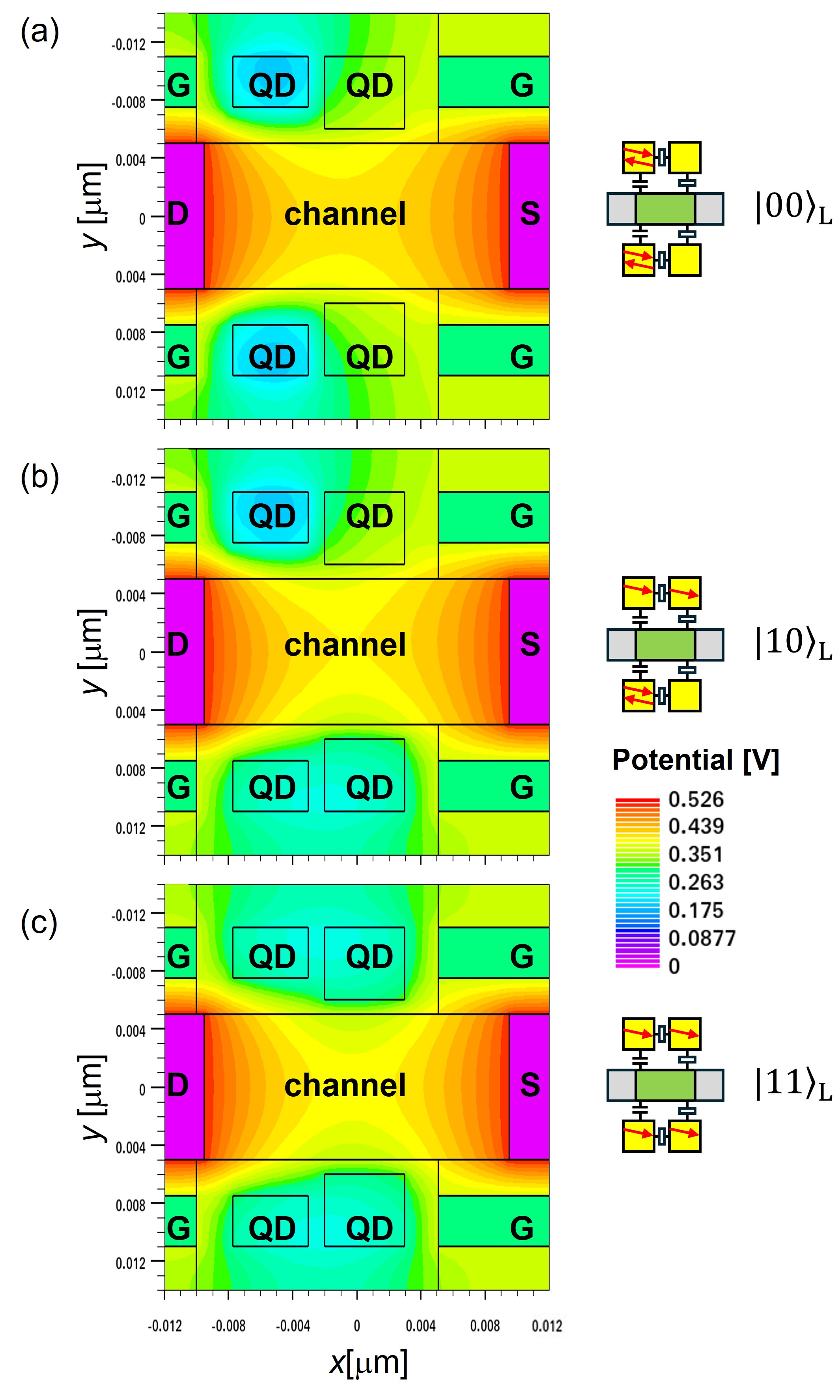}
\caption{Electric potential profile of the $x$-$y$ cross-section at 
$V_{\rm g}=0.3$ V and $V_{\rm D}=0.0001$ V, for the medium sized 
quantum dot (5 nm):
(a) $|00\rangle_L$ state (b) $|10\rangle_L$ state (c) $|11\rangle_L$ state.
}
\label{fig_potential}
\end{figure}

\begin{figure}
\centering
\includegraphics[width=8.0cm]{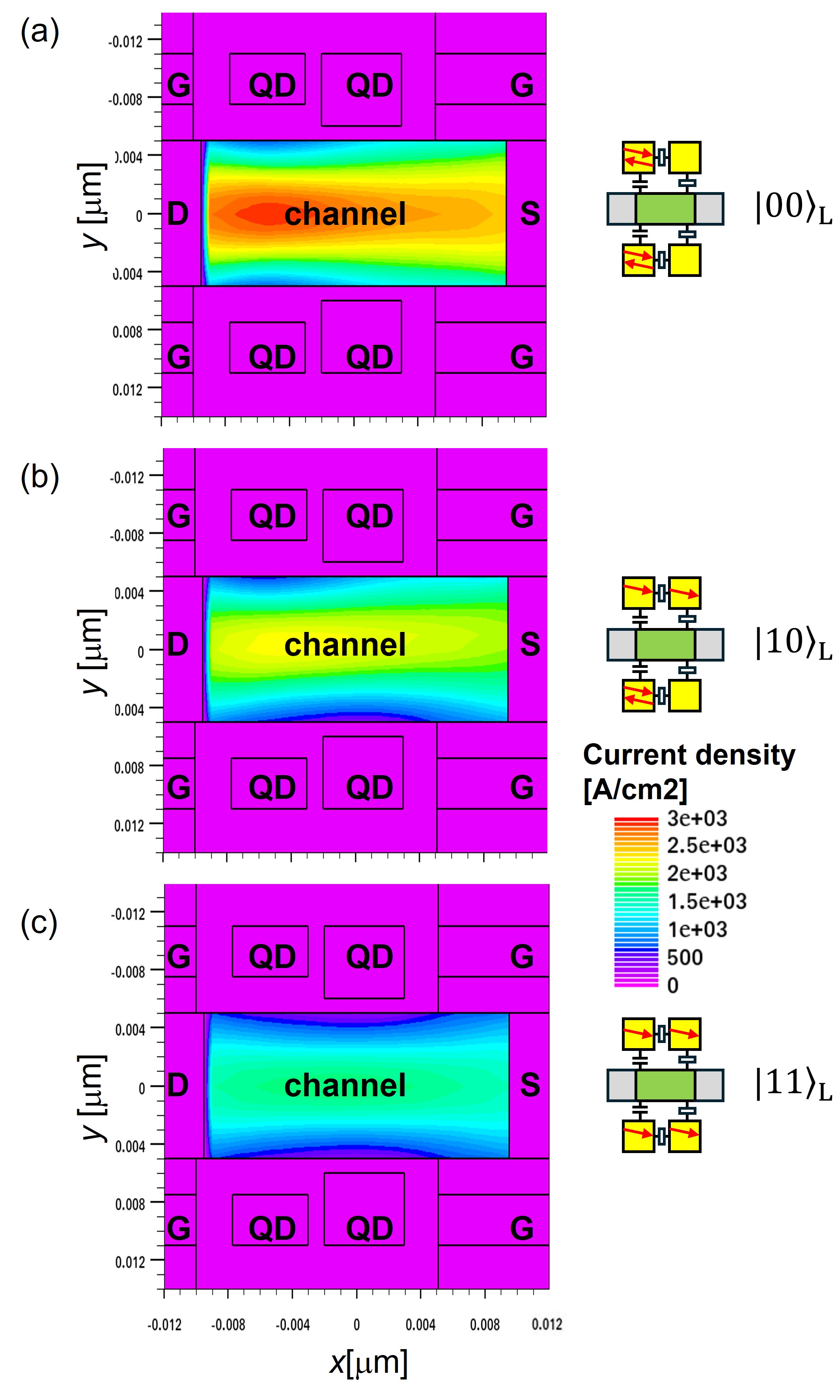}
\caption{Current density profile of the $x$-$y$ cross-section at
$V_{\rm g}=0.3$ and $V_{\rm D}=0.001$V, for the medium sized 
quantum dot (5 nm):
(a) $|00\rangle_L$ state (b) $|10\rangle_L$ state (c) $|11\rangle_L$ state.
}
\label{fig_cross_current}
\end{figure}

\section{Results of TCAD simulations}\label{sec:TCAD}
\subsection{$I$-$V$ characteristics}
To construct a logical qubit (see Fig. 1), two electrons must be inserted into the coupled quantum dots. 
This initialization process is achieved by applying a large \( V_{\text{cg}} \), 
which allows electrons to enter the coupled quantum dots from the GAA channel. 
The presence of electrons in quantum dots is confirmed by measuring the channel currents. 
Figure \ref{fig_plane_pre} illustrates the currents in the following three cases: 
(i) No electrons present. 
(ii) Electrons only in the left quantum dot. 
(iii) Two electrons present in both quantum dots. 
Similar results were observed for other charge distributions between cases (i) and (iii) (data not shown).
As the number of charges increases, the current is suppressed for various charge distributions.
This is crucial for initializing the system, as it helps verify that the desired configuration has been established.

Next, we show the numerical results for the logical states, assuming that two electrons exist in the coupled quantum dots.
Figure~\ref{fig_plane_distributions} presents the $I_{\rm D}$-$V_{\rm D}$ and $I_{\rm D}$-$V_{\rm G}$ characteristics for three different sizes of quantum dots, 
and various charge distributions $|00\rangle_L$, $|10\rangle_L$, and $|11\rangle_L$. 
The state $|0\rangle_L$ indicates that two electrons are present in the same quantum dot, 
whereas $|1\rangle_L$ indicates that each quantum dot contains one electron.
Consequently, we can define the two-qubit states $|00\rangle_L$ $|10\rangle_L$, and $|11\rangle_L$ according to these diverse charge distributions. 
The current for state $|01\rangle_L$ is identical to that for $|10\rangle_L$. 
As shown in Fig.~\ref{fig_plane_distributions}, the current varies based on the logical states, owing to the differences in the charge distributions. 
It decreases in the order of $|00\rangle_L$, $|10\rangle_L$, and $|11\rangle_L$. 
This suggests that the localized charge configuration $|00\rangle_L$ results in the least current suppression, whereas the widely distributed charge configuration $|11\rangle_L$ results in the greatest suppression. 
The latter occurs because a broader charge distribution interferes with the effect of the gate electrodes over a larger area.
These calculations confirm that the current is influenced by the charge distribution within the double quantum dots.

For small quantum dots (2.5 nm), the area covered by two quantum dots was 5 nm, which is equivalent to that of a single medium-sized quantum dot (5 nm). 
Consequently, the current-voltage characteristics shown in Figs.~\ref{fig_plane_distributions} (a) and (b) are greater than those of the medium-sized quantum dots in Figs.~\ref{fig_plane_distributions} (c) and (d).
From Table 1, in the case of large quantum dots (10 nm), two quantum dots with a total length of 20 nm almost cover the channel area, which is 22 nm long, 
significantly reducing the gate area without quantum dots. 
However, the current produced by large quantum dots (Figs.~\ref{fig_plane_distributions} (e) and (f)) is comparable to that for medium-sized quantum dots (5 nm) 
(Figs.~\ref{fig_plane_distributions} (c) and (d)).
Distinct charge distributions are observed in all the three cases. 
Because variations in quantum dot size are unavoidable during fabrication, 
quantum dots of different sizes are inevitable. 
Our results indicate that the currents reflect the different qubit states associated with the varying quantum dot sizes.

\subsection{Potential and Current Density Profile}
In this section, to demonstrate the effects of different charge distributions on the electronic profile of the GAA structure, 
we present the numerical results for the spatial potential and current density profiles. 
Figure~\ref{fig_potential} illustrates the potential distribution in the $x$--$y$ cross-section at the channel center ($z=0$). 
Importantly, the operating voltage range of qubits is significantly lower than that of conventional transistors. 
A large voltage drop near the qubits can severely degrade their coherence; thus, it is crucial to monitor the voltage drops in their vicinity.
In Fig.~\ref{fig_potential}, even in the low $V_D$ and $V_G$ regions, 
different qubit states produce distinct potential profiles. 
For instance, examining the channel region reveals that the electric potential around $x=0.0$ for state (a) is higher than that for state (c). 
Conversely, state (c) exhibits a significant potential drop. 
This difference arises because gate controllability is diminished owing to the broad charge distribution in state (c). 
The electric potential for state (b) lies between those of states (a) and (c).

In the present configuration, the quantum dots are separated from the GAA channel by a 1-nm SiO${}_2$ insulator. 
The estimated breakdown voltage for 1-nm SiO${}_2$ typically ranges from 1.0 to 1.2 V, which corresponds to an electric field of approximately 10 MV/cm~\cite{McPherson}. 
From this perspective, the voltage drop across SiO${}_2$ in Fig.\ref{fig_potential} is $\leq$ 0.1 V, preventing the breakdown of the SiO${}_2$.

Figure~\ref{fig_cross_current} shows the distribution of the current density in the $x$--$y$ cross-section at the channel center (at $z=0$). 
The current density distribution varies according to the charge distribution, 
which is influenced by the qubit states, similar to the patterns shown in Fig.\ref{fig_potential}. 
Initially, it was inferred that this $|00\rangle_L$ state considerably obstructs the current near $x = -0.006 \, \mu \text{m}$. 
However, our calculations indicate that this effect is weak. 
Instead, we found that the wide-range disruption caused by the control gates, 
as in  $|11\rangle_L$, has a dominant influence on the current distribution.

In a qubit system, the number of electrons should be countable,
and the operating voltage is significantly lower than that of conventional transistor systems. 
Therefore, when a quantum system is connected to a conventional transistor circuit, 
the current range of the former must be significantly lower than that of the latter.
For example, an $I=1$ nA current includes  
$I/e = 10^{-9}/1.6\times 10^{-19}\approx 10^{10} \text{ electrons/s}
$ (where $e$ denote the electron charge).  
When the current is measured at approximately 100 MHz, 
the target number of electrons, which is countable, is calculated as follows:  
$10^{10}/10^8 = 100$.
We use a current of $\leq$ 1 nA as the guideline for the quantum system.

\section{Results of circuit simulations}~\label{sec:SPICE}
The differences in the current characteristics among the various qubit states are small compared to the output of conventional transistors. 
Therefore, it is necessary to amplify the output signals. 
We can incorporate TCAD data into circuit simulations using Verilog-A. 
In this study, we conducted circuit simulations based on TCAD data, similar to the approach used in \cite{TanaAPL}.

The Berkeley Short-Channel IGFET Model (BSIM) version 106, level 72, was utilized in Silvaco SmartSpice. 
In \cite{TanaJAP2023, TanaJAP2025}, the typical operational temperature of qubits is 100 mK. 
Because the transistors and qubits are arranged side-by-side, 
transistor circuits connected to the qubit system operate at similarly low temperatures. 
However, the circuit simulations in Silvaco SmartSpice do not converge properly below approximately 10 K. 
Therefore, the circuit simulations presented below were conducted at \(T = 10\)~K.

\begin{figure}
\centering
\includegraphics[width=8.0cm]{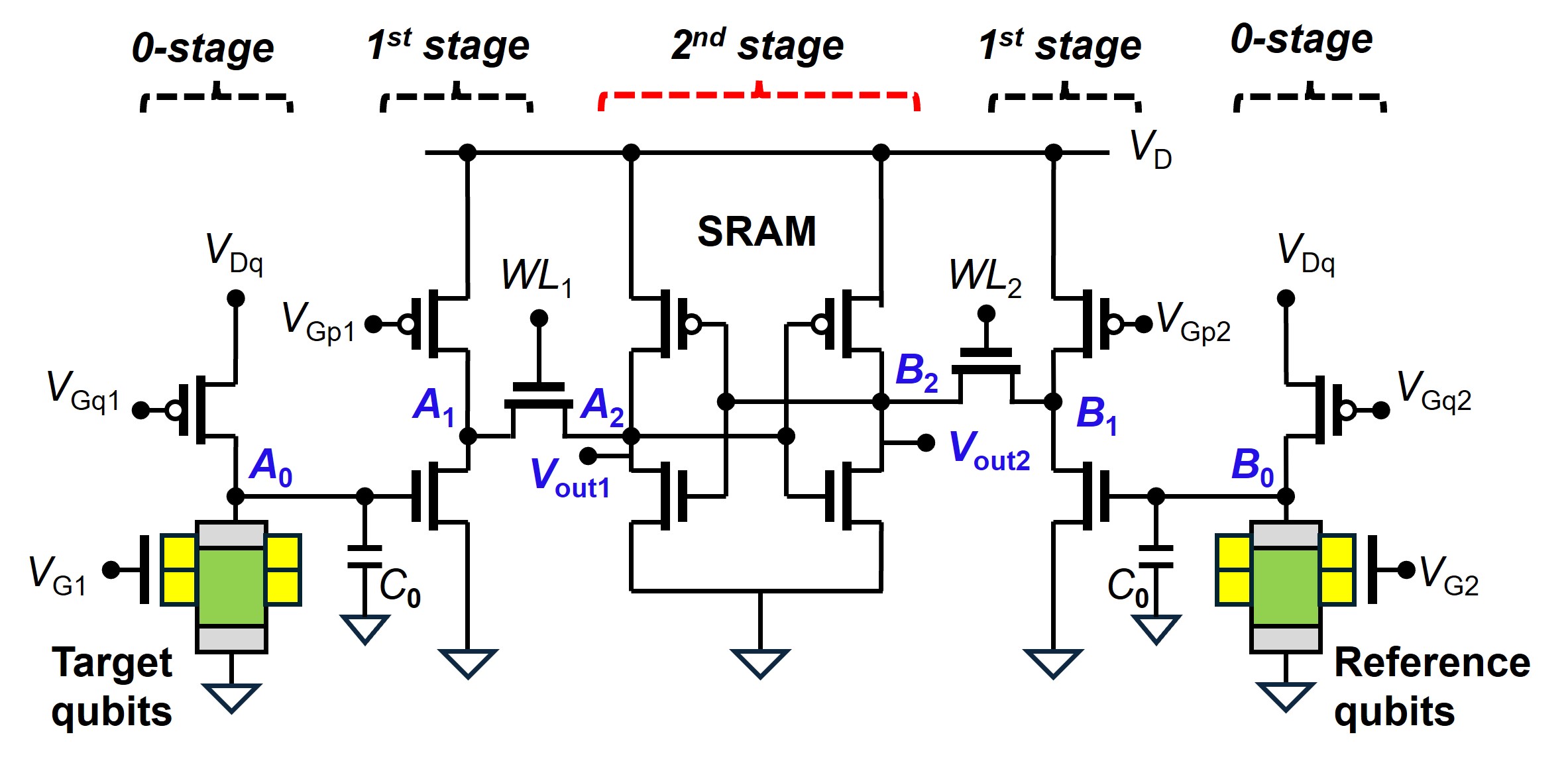}
\caption{
Qubit readout circuit based on the comparison of two GAA outputs.
Left and right GAA transistors are the target and reference qubits, respectively. 
The $I$-$V$ data obtained from the TCAD simulations are input to the 
SPICE circuit calculations through Verilog-A.
The gate length of PMOS and NMOS transistors are 28 and 14nm finfets, respectively.
The fin height of the PMOS transistor is  
20, 10, and 20nm for the 0-,1-,2-stages, respectively.
Those of the NMOS transistor are 20, 35, and 35nm, respectively.
The capacitance is $C_0=0.05$pF.
We also add capacitances 0.2pF to $V_{\rm out1}$ and $V_{\rm out2}$.
}
\label{fig_sram1}
\end{figure}

Figure~\ref{fig_sram1} shows the proposed readout circuit. 
The outputs of the target qubits are compared with those of the reference qubits whose states are known. 
Weak signals from the qubits are amplified in three stages, progressing from $A_0$ ($B_0$) to $A_2$ ($B_2$). 
The amplified signal at $A_2$ is then compared to the signal at $B_2$ based on the static random access memory (SRAM) 
configured as a sense amplifier, 
located at the center of Fig.~\ref{fig_sram1}. 
These signals are latched in the second stage, producing two outputs: $V_{\rm out1}$ (equal to \(A_2\)) and $V_{\rm out2}$  (equal to \(B_2\)). 
The qubit states are subsequently detected as digital signals of 0 or 1 using conventional CMOS circuits (not shown). 
This circuit was originally proposed in \cite{TanaAPL}. 
In this design, the CMOS transistors located below $A_1$ and $B_1$ are used instead of the single electron transistors mentioned in \cite{TanaAPL}.

The following two main points must be addressed.
The first point is to confirm the gradual increase in the qubit output voltage from \(A_0\) and \(B_0\) to \(A_2\) and \(B_2\). 
As mentioned previously, 
the current flowing through the GAA transistor at the 0-stage must be limited to the order of nA to reduce backaction on the qubits. 
Therefore, $V_{\rm Dq}$ should be sufficiently small such that the voltages at \(A_0\) and \(B_0\) are on the order of meV. 
However, the voltages at \(A_2\) and \(B_2\) should be approximately $V_{\rm D}/2$ to facilitate comparison between the target and reference qubits.
The second point is to manage the backaction during the latching process. 
Outputs $V_{\rm out1}$  and $V_{\rm out2}$ are ultimately latched to either zero or $V_{\rm D}$, starting from \(V_{\rm D}/2\). 
During this latching process, a significant amount of current flows directly from the power supply \(V_{\rm D}\) to the ground, 
because both PMOS and NMOS transistors are turned on simultaneously for a brief period. 
This shoot-through current can affect the qubits, representing a backaction resulting from the measurement.

To address the first point, we controlled the voltages \(A_0\)--\(A_2\) and \(B_0\)--\(B_2\) by selecting different sizes of the PMOS and NMOS transistors. 
The gate lengths of the PMOS transistors are chosen to be longer than those of the NMOS transistors. 
By making NMOS transistors more conductive than PMOS transistors, the voltage levels of NMOS transistors are reduced.
For the second point, we control the wordline voltages ($V_{\rm WL}$) dynamically. 
A appropriate voltage profile is generated to control the dynamic response during the latching process.

\begin{figure}
\centering
\includegraphics[width=8.5cm]{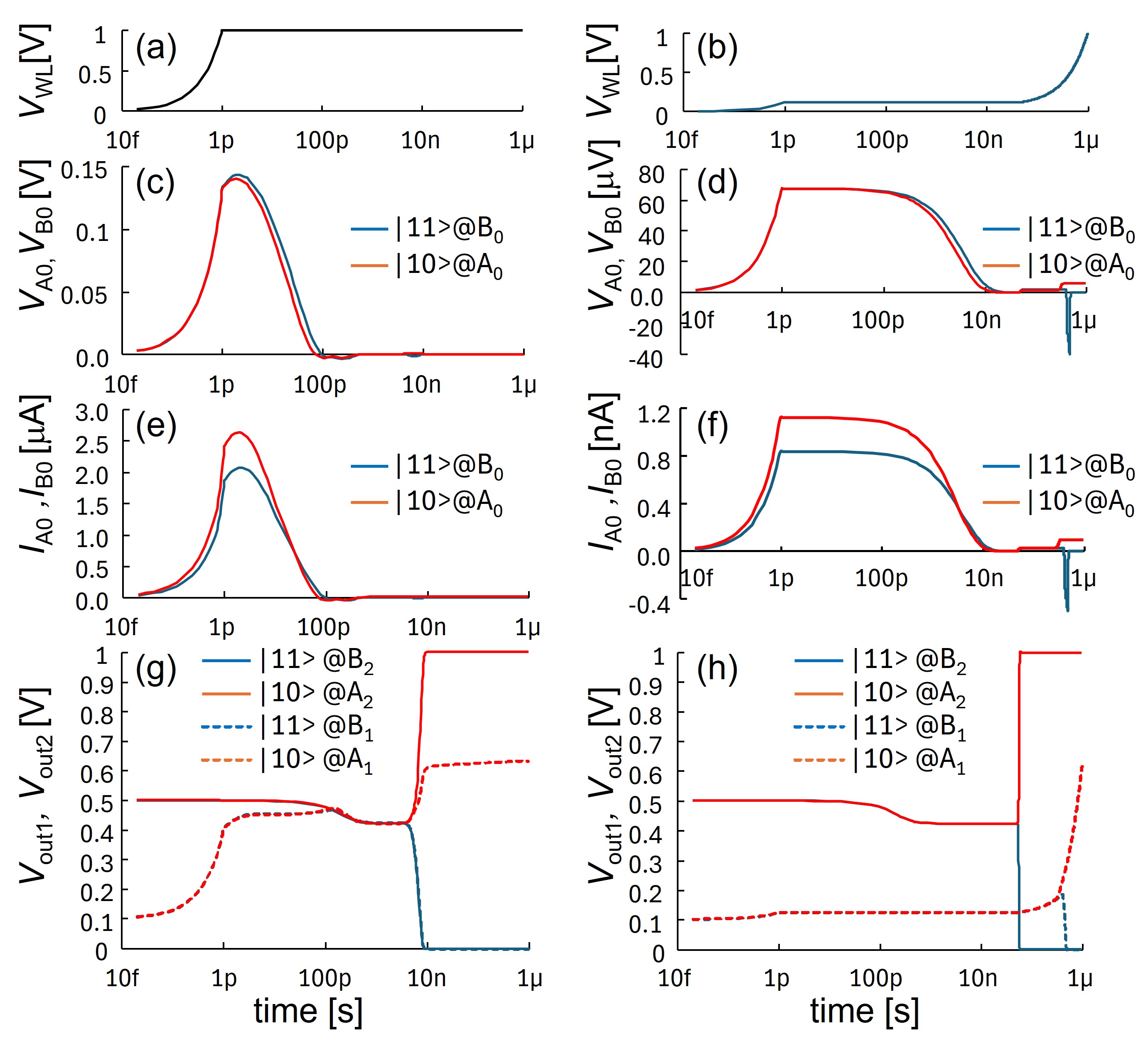}
\caption{
Time-dependent circuit behaviors for two types of 
voltage control of $V_{WL}$. 
(a) Straightforward switching of $V_{WL_i}$. 
(b) Modified switching of $V_{WL_i}$ ($i=1,2$).
(c,e,g) Response of the circuit for the switching in (a).
(d,f,h) Response of the circuit for the switching in (b).
$T=10$K. The target qubit is in the $|11\ra_L$ state, and 
the reference qubit is in the $|01\ra_L$ state.
}
\label{fig_sram2}
\end{figure}

To provide a comparison, we show how the circuit operates during the conventional switching process.
Figures~\ref{fig_sram2}(a), (c), (e), and (g) show the circuit node profiles when the wordline voltage \(V_{\rm WL}\) is activated, 
ramping to \(V_{\rm D}\) within 1 ps (Fig.~\ref{fig_sram2}(a)). 
The qubit states \(|10\rangle_L\) and \(|11\rangle_L\) are compared. 
Figures~\ref{fig_sram2}(b) and (c) show the voltages and currents at nodes \(A_0\) and \(B_0\). 
A considerable amount of current flows, significantly changing the qubit voltage. 
Those large voltage and current fluctuations are undesirable for qubits, 
as mentioned in the previous section.
They can induce substantial voltage drops and generate hot electrons, destroying the qubits' coherent states. 

\begin{figure}
\centering
\includegraphics[width=8.0cm]{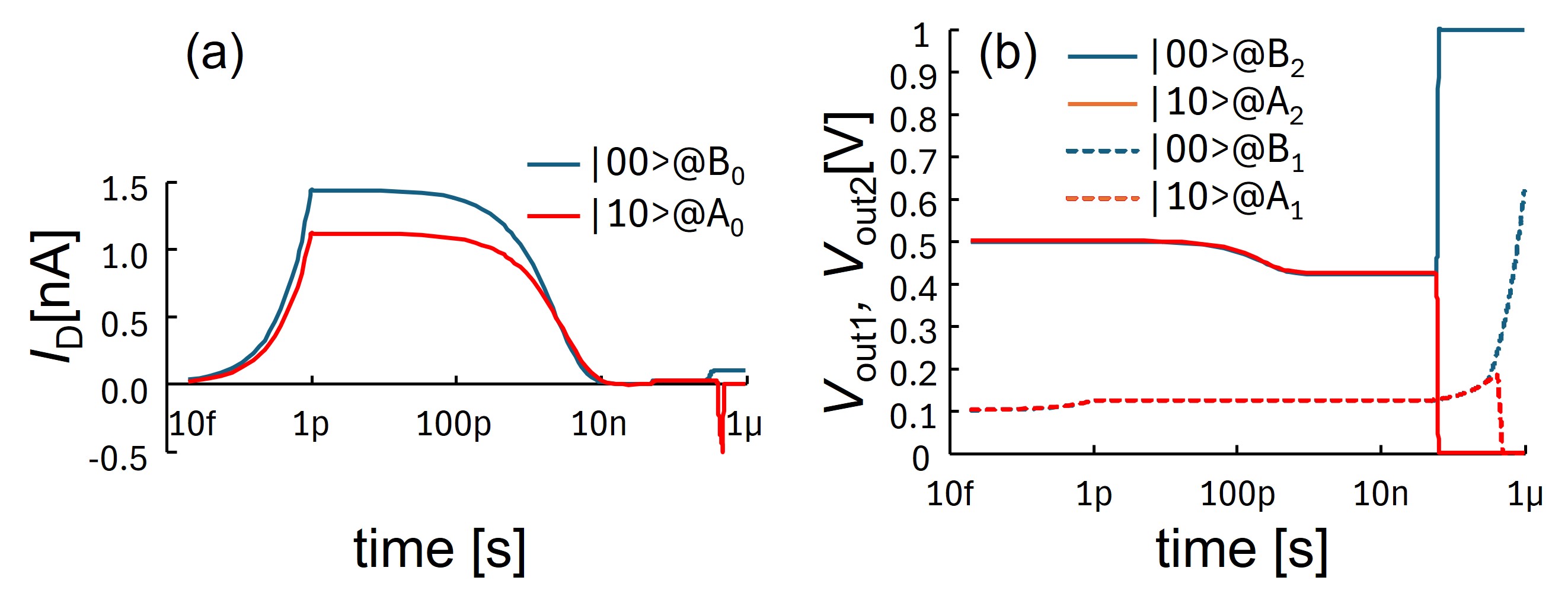}
\caption{
Time-dependent circuit behaviors for two types of 
voltage control of $V_{WL}$ of Fig.~\ref{fig_sram2}(b). 
The target qubit is in the $|00\ra_L$ state, and 
the reference qubit is in the $|01\ra_L$ state.
}
\label{fig_sram3}
\end{figure}

To mitigate the backaction by the measurement, it is advisable to maintain nodes \(A_0\) and \(B_0\) in lower-voltage states. 
We propose two methods for achieving this:
(i). The switching profile of the wordline voltage begins at a low voltage such as 0.12 V.
(ii). The small capacitors \(C_0\) are attached to nodes \(A_0\) and \(B_0\). 
Figures~\ref{fig_sram2}(b), (d), (f), and (h) present the adjusted wordline voltage as well as the voltages and currents at \(A_0\) and \(B_0\), 
along with $V_{\rm out1}$ and $V_{\rm out2}$. 
In Fig.~\ref{fig_sram2}(b), the wordline voltage is suppressed for approximately 20 ns before increasing linearly to \(V_{\rm D}\). 
In Figs.~\ref{fig_sram2}(d) and (f), the voltage and current are suppressed, as expected. 
Figure~\ref{fig_sram2}(h) shows that the qubit states can be output as either zero or \(V_{\rm D}\), 
although the latching time is slightly delayed compared to that in Fig.~\ref{fig_sram2}(g). 
Figure~\ref{fig_sram3} shows results similar to those in Fig.~\ref{fig_sram2} for the qubit states \(|01\rangle_L\) and \(|11\rangle_L\).
Thus, through careful design of the transistor size and dynamic control of the wordlines, 
we can effectively detect qubit states while minimizing the backaction of the measurement. 
A more detailed assessment of the effects of backaction, 
such as its impact on coherence time, lies beyond the scope of this study and is a topic for future research.

\section{Discussions}~\label{sec:discussions}
The readout mechanisms described in \cite{TanaJAP2023,TanaJAP2025} utilize the change 
of the resonant energy levels that match the qubit states.
Thus, the channel part should detect the resonant energy--levels of a single quantum dot.
Here, instead of directly detecting the energy-levels, 
the charge distributions of the coupled quantum dots are detected, 
which is more suitable for conventional transistors.
Our structure is intended to be extended from the conventional nanosheet transistors.
This is because the change in the established fabrication structure 
should be as small as possible to avoid additional fabrication difficulties and limit costs.
We assume that the quantum dots are embedded within the gate structures.
While the current proposal is intended to be closer to the nanosheet structure than \cite{TanaJAP2023,TanaJAP2025}, 
it is obvious that new complicated fabrication processes to create quantum dots are necessary. 
This still remains a challenge for future research.

In our SPICE circuit simulations, we assumed there were no variations in the CMOS transistors. 
The variation in FinFET devices can be as high as 50 meV~\cite{Wang,Zhang2}, 
which exceeds the signal range of qubits. 
However, provided that the initial calibration is conducted properly, conventional transistors can be directly connected to the qubit system by adjusting the appropriate voltages. 
This is feasible because the variations in transistors are consistent and serve as the unique fingerprint of each device~\cite{Holcomb}. 
A detailed study of the effects of these variations is a focus for our near future.
In \cite{TanaAIP}, we discussed the effects of noise and traps on the compact spin qubit system. 
The influence of noise and traps on the current structure will be also addressed in the future.

\section{Conclusions}~\label{sec:conclusions}
We propose a theoretical spin-qubit system based on a GAA transistor that integrates spin qubits. 
In this system, GAA transistors serve as both measurement devices and  means for interaction between qubits. 
Each qubit consists of two quantum dots, with a logical qubit state defined by the parallel and antiparallel spins within these quantum dots.
The current flowing through the GAA channel is computed using TCAD simulations. 
The difference in qubit states can be determined from the GAA channel current. 

Qubit state determination is facilitated by SRAM connected to the qubit system.
To prevent degradation of the qubit state, we designed a readout circuit to apply voltages to qubits that are as low as possible. 
A more precise estimation of the effects of the measurement on qubits will be explored in future work. 
For instance, this can be achieved by solving the density matrix equations of the qubit during detection. 
This remains a challenge for our research.

\appendix

\section{Type-B Structure}~\label{sec:Appendix_A}
\begin{figure}
\centering
\includegraphics[width=8.8cm]{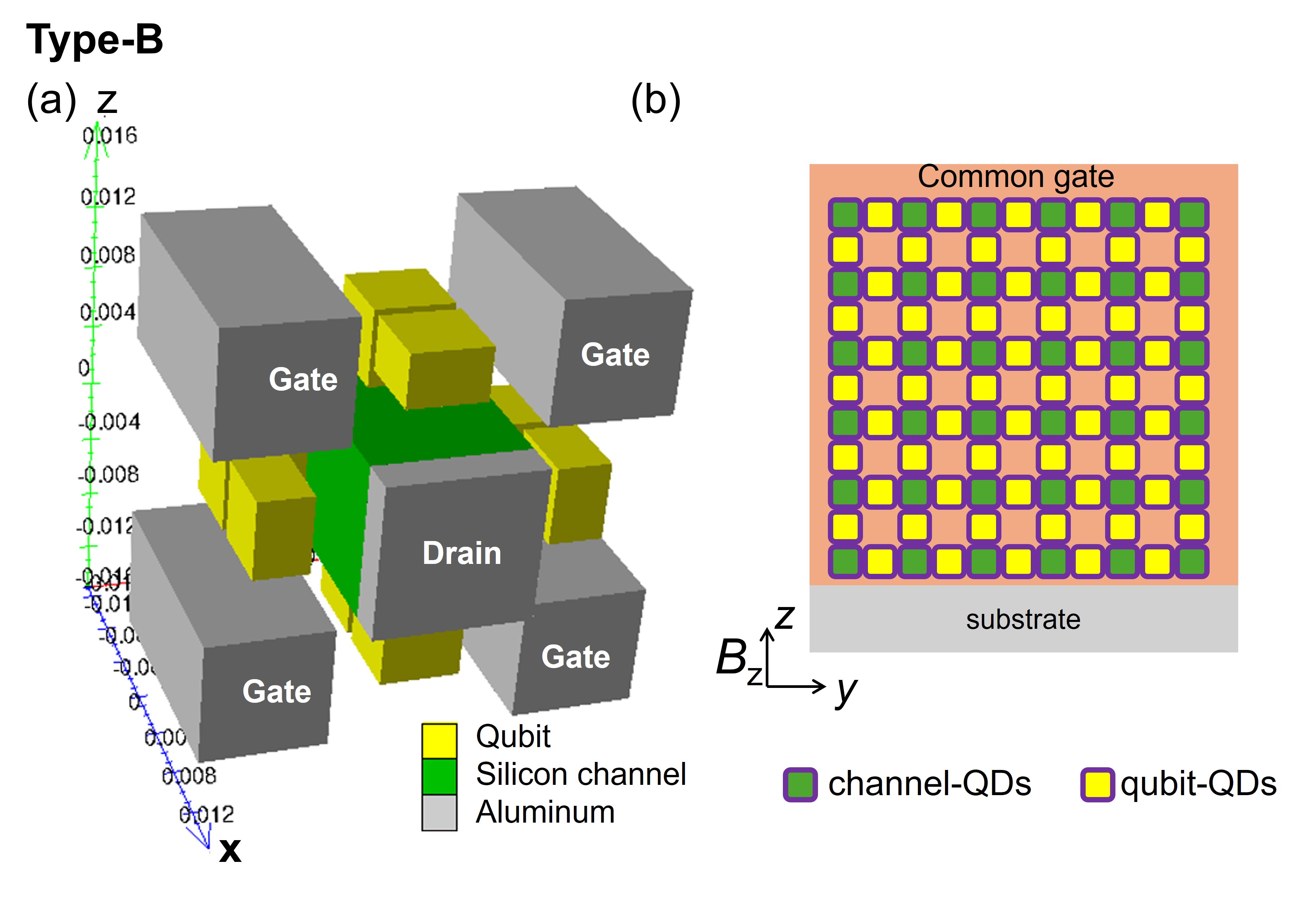}
\caption{
We can arrange the qubits and GAA transistor differently from type-A.
(a) A single unit. (b) 2D qubit array of (a). 
In this configuration (type-B), the GAA structure (green) is surrounded by four qubits (yellow) on four sides ((a)).
The qubits and GAA transistor are surrounded by the common gate via SiO${}_2$, 
which is not shown in the figure.  
}
\label{fig_typeB}
\end{figure}

Figure~\ref{fig_typeB} (a) presents a stacking structure that differs from that of type-A, 
as shown in Fig.~\ref{fig3}. 
In the type-B configuration, the GAA transistor is surrounded by four coupled quantum dots. 
The current generated by the GAA transistor reflects more complex charge distributions 
than that generated by the type-A configuration. 
In addition, during the interaction process, the four qubits share a single GAA transistor, 
which must manage the control of the qubits from different directions. 
Thus, more operational steps are necessary compared to the type-A configuration.
As illustrated in Fig.~\ref{fig_typeB}(b), the common gate covers the qubit more extensively than in type-A (see Fig.~\ref{fig3}(b)). 
This suggests that the controllability of the common gate is stronger in type-B than that in type-A.

\begin{figure*}
\centering
\includegraphics[width=15.8cm]{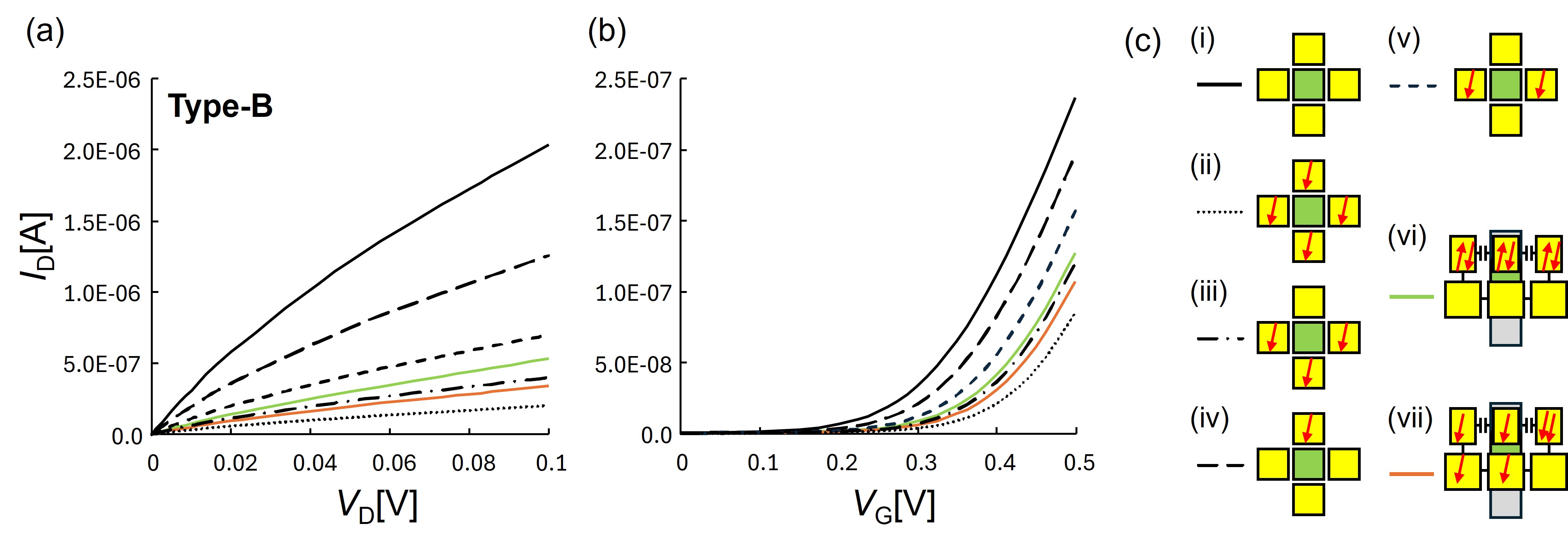}
\caption{(a) $I_D$-$V_D$ characteristics and (b) $I_D$-$V_G$ characteristics
of the type-B structure for various charge distributions.
(c) Electron distributions in (a) and (b).
The quantum dot size is 5 nm, similar to the "standard" pattern in Table 1.
}
\label{figcubic}
\end{figure*}

Figure~\ref{figcubic} shows examples of the current characteristics of the type-B structure. 
Similar to the observations presented in Fig.~\ref{fig_plane_distributions}, 
the differences in the current due to the different charge distributions are noticeable. 
Consequently, the detected current reflects the charge distributions in four distinct directions in the type-B configuration. 
Determining which configuration---type-A or type-B--- is superior for the qubit system remains an open question for future research.


\section{Qubit operations}~\label{sec:Appendix_B}
Loss and DiVincenzo proposed a CNOT gate between two qubits 
using single-qubit rotations and Heisenberg interactions to form a $\sqrt{\rm SWAP}$ gate~\cite{Loss}.
However, the formation of a CNOT gate using single-qubit rotations and Heisenberg interactions 
is engineeringly problematic.
For example, Srinivasa {\it et al.} theoretically investigated the mediated electron interaction~\cite{Srinivasa}, 
and found that the magnitude of the qubit-qubit coupling was $J \sim 0.2$ neV, 
which results in an extremely slow operation time of approximately $\tau_{\rm gate} \sim 5 \mu$s. 
Meh {\it et al.} explored a mediated electron system using singlet-triplet states~\cite{DiVincenzo}. 
In this case, the coupling between the two qubits is proportional to $t^4$, 
where $t$ denotes the tunneling coupling between the two quantum dots. 
Therefore, from an engineering perspective, achieving precise control of mathematical CNOT gate formation seems difficult. 
CNOT gates have been realized experimentally using resonant energy levels~\cite{Zajac,TaruchaCNOT}. 
In this context, to realize CNOT gates using our proposed method,
it is sufficient to analyze the eigenvalues of the Hamiltonian Eq.(\ref{Hamiltonian}) 
and determine the region where four energy-levels are separated from each other.

The qubit operations are performed using interactions between electron spins.
The spin-spin interaction is mediated by the Heisenberg exchange interaction 
via the direct tunneling of electrons through the thin tunneling oxide.
The interaction between neighboring two quantum dots in the same logical qubits is always switched on.
However, the interaction between quantum dots via the GAA channel 
is switched on only when the Fermi energy of the GAA channel is close to that of the quantum dots,  
as discussed in \cite{TanaAIP}. 
Figure~\ref{fig_concept}(a) illustrates the interactions between the four qubits, 
where the Hamiltonian is given by:
\begin{equation}
H=J_{12}{\mathbf S}_{1}\cdot{\mathbf S}_{2} + J_{23}{\mathbf S}_{2}\cdot{\mathbf S}_{3} +J_{34}{\mathbf S}_{3}\cdot{\mathbf S}_{4} +
\sum_{i=1}^4 g\mu_B {\mathbf S}_{i}\cdot {\mathbf B}_i,
\label{Hamiltonian}
\end{equation}
where ${\mathbf S}_{i}$, $J_{ij}$, and ${\mathbf B}_i$ denote the spin vector, 
the coupling between two spin qubits, and the applied magnetic field, 
respectively. 
$J_{12}$ and $J_{34}$ originate from the direct tunneling between the nearest quantum dots.
$J_{23}$ is mediated by the channel electrons (see Section~\ref{sec:Appendix_C}).
$g$ and $\mu_B$ denote the $g$-factor and Bohr magneton, respectively.
Hereinafter, we assume $g=2$.
We also assume that the parameter regions are given by
\begin{equation}
 J_{23} \ll J_{12}, J_{34} \ll |{\mathbf B}_i|.
\end{equation}

As shown in Fig.~1, we consider a logical qubit state comprising $|\dna\upa\ra$ and $|\dna\dna\ra$ states.
Then, we prepare four states---  
$|\dna\dna\ra|\dna\dna\ra$, $|\dna\dna\ra|\upa\dna\ra$, 
$|\upa\dna\ra|\dna\dna\ra$, and $|\upa\dna\ra|\upa\dna\ra$---
for QD${}_1$-QD${}_4$ in Fig.~\ref{fig_concept}.
First, we consider the case where $J_{i,i+1}=0$ ($i=1,2,3$) under a uniform magnetic field.
The lowest energy level corresponds to the $|\dna\dna\dna\dna\ra$ state.
However, the excited energy levels are superpositions of 
$|\dna\dna\ra|\dna\upa\ra$, $|\dna\dna\ra|\upa\dna\ra$, 
$|\dna\upa\ra|\dna\dna\ra$, and $|\upa\dna\ra|\dna\dna\ra$ 
(one spin is flipped).
Thus, no $|\upa\dna\ra|\upa\dna\ra$ state exists.
In the case of no magnetic field, 
the ground state is a superposition of $|\upa\dna\ra|\upa\dna\ra$, $|\upa\dna\ra|\dna\upa\ra$, 
$|\dna\upa\ra|\upa\dna\ra$, and $|\dna\upa\ra|\dna\upa\ra$. 
To choose the four states  
$|\dna\dna\ra|\dna\dna\ra$, $|\dna\dna\ra|\upa\dna\ra$, 
$|\upa\dna\ra|\dna\dna\ra$, and $|\upa\dna\ra|\upa\dna\ra$, 
as the lowest four states, the magnetic field should be limited.
In addition, a nonuniform magnetic field must be applied with a gradient in the $x$ and $y$ directions.

We solve the basic eigenvalue problem in Eq.(\ref{Hamiltonian}) under a nonuniform magnetic field.
For the $|\upa\upa\ra$, $|\upa\dna\ra$, $|\dna\upa\ra$ and $|\dna\dna\ra$ basis,
the Hamiltonian under a nonuniform magnetic field is expressed as
\begin{equation}
H
=\left(
\begin{array}{cccc}
   E_z+J/4  &  \mu_B B_-^R   &\mu_B B_-^L   & 0 \\
\mu_B B_+^R &  -\delta E_z-J/4 &  J/2           & \mu_B B_-^L  \\
\mu_B B_+^L &  J/2             & \delta E_z-J/4 & \mu_B B_-^R  \\
        0   &  \mu_B B_+^L   & \mu_B B_+^R  & -E_z+J/4 
\end{array}
\right),
\label{Hamiltonian1}
\end{equation}
where $B_\pm^\alpha \equiv  B_x^\alpha \pm iB_y^\alpha$, 
$E_z=B_z^h+(B_z^L+B_z^R)/2$, and $\delta E_z=(B_z^R-B_z^R)/2$.

Figures~\ref{fig_concept} (b) presents an example of the eigenvalues in Eq.(\ref{Hamiltonian1}).
The eigenvalues depend on the parameters.
The lowest four states are separated from the excited states.
Table 2 presents the eigenvalues and eigenfunctions of the four lowest states.
For this case, we select the four lowest energy levels: 
$|\Psi_{0}\ra \sim |\Psi_{3}\ra$.
These states correspond to superposition of 
the $|\upa\upa\ra$, $|\upa\dna\ra$, $|\dna\upa\ra$, and $|\dna\dna\ra$ states.

\begin{table*}
\begin{center}
\caption{\textbf{Eigenvalues for Eq.(\ref{Hamiltonian1})}}
\label{table2}
\setlength{\tabcolsep}{3pt}
\begin{tabular}{ c|l}
\hline
Eigenenergy [meV] &
Eigenfunction  \\
\hline  
$E_{0}=$ $-0.250$ & $|\Psi_{0}\ra\approx 0.990|\dna\dna\dna\dna\ra$ \\ 
$E_{1}=$ $-0.188$ & $|\Psi_{1}\ra\approx 
   0.657|\dna\dna\upa\dna\ra -0.711|\dna\upa\dna\dna\ra$ 
 $+0.176|\dna\upa\dna\dna\ra -0.150|\dna\dna\dna\upa\ra$ \\
$E_{2}=$ $-0.183$ & $|\Psi_{2}\ra\approx 
  -0.166|\dna\dna\upa\dna\ra +0.162|\dna\upa\dna\dna\ra$  
 $-0.708|\dna\upa\dna\dna\ra -0.660|\dna\dna\dna\upa\ra$\\
$E_{3}=$ $-0.120$ & $|\Psi_{3}\ra\approx 
 -0.504|\upa\dna\upa\dna\ra$ $+0.470|\upa\dna\dna\upa\ra$  
$+0.520|\dna\upa\upa\dna\ra  -0.504|\dna\upa\dna\upa\ra$ \\
\hline
\multicolumn{2}{p{251pt}}{
Eigenvalues and eigenfunctions for the lowest four states at $B_z=0.07 $meV. }\\
\end{tabular}
\end{center}
\end{table*}
%
%
%
%

\section{Estimation of Qubit-Qubit Interaction via GAA Channel}~\label{sec:Appendix_C}
\begin{figure*}
\centering
\includegraphics[width=15.8cm]{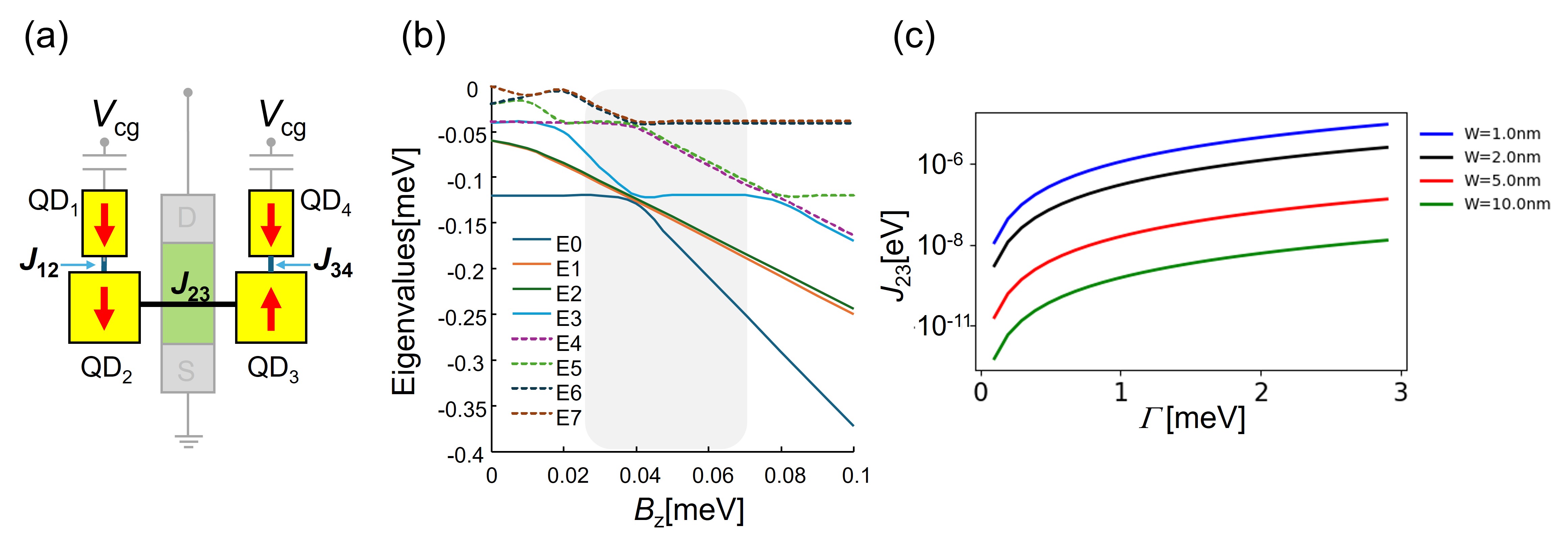}
\caption{
The qubit operation is based on the Heisenberg coupling between quantum dots.
(a) Qubit-qubit interaction between the left quantum dot (QD2) and right one (QD3) 
is mediated by the channel electron at $V_{\rm D}=0$.
This is called RRKY interaction~\cite{Ruderman,Kasuya,Yosida,Rikitake}.
(b) Example of the eigenvalues obtained by diagonalizing the 
Hamiltonian Eq.(\ref{Hamiltonian1}).
$B_x=0.01$ meV, $J_{12}=J_{34}=0.02$ meV, and $J_{23}=0.001$ meV. 
For $dB_z=0.03$ meV, 
$B_z^{(1)}=B_z(1.0+2dB_z)$ meV, $B_z^{(2)}=B_z(1.0+dB_z)$ meV, $B_z^{(3)}=B_z$ meV, 
and $B_z^{(4)}=B_z(1.0+dB_z)$ meV.
The shaded area indicates the ideal parameter region for realizing Fig.1.
(c) Magnitude of the RKKY interaction as a function of the tunneling coupling 
for various channel width.
The electron density is $n_3=10^{19}$/cm$^3$, $E_m=25$meV, and $E_F=84.6$meV.
}
\label{fig_concept}
\end{figure*}

The interaction between distant qubits is mediated by channel electrons 
when \( V_{\rm D} = 0 \) (depicted as \( J_{23} \) in Fig.~\ref{fig_concept}(a)). 
This interaction is known as the Ruderman-Kittel-Kasuya-Yosida (RKKY) interaction. 
It arises from the \( s \)--\( d \) coupling between a localized electron and a channel electron, 
which mediates the interaction between distant localized electrons~\cite{Ruderman,Kasuya,Yosida}. 
Experiments~\cite{Sasaki,Marcus,Croot} indicated that the RKKY interaction is an important mechanism for facilitating interactions between spin qubits.

The magnitude of the RKKY interaction for quantum dots was estimated as\cite{Rikitake,TanaAIP}: 
\begin{equation}
J_{23} = 8\alpha_3 E_F F'_3(2k_FW),
\end{equation}
where \(\alpha_3\) is defined as follows: 
$
\alpha_3 = \frac{m^2 k_F^2 J_{sd}^2}{16\pi^3\hbar^3},
$
and $F'_3(x) = (x \cos x - \sin x)/x^4$. 
Here, \(E_F\) denotes the Fermi energy, \(W\) denotes the channel width, 
and \(k_F = (3\pi^2 n_e)^{1/3}\) denotes the Fermi wave number, with \(n_e\) being the electron density. 
Following the approach outlined in \cite{Schrieffer}, 
we assume that the \(s\)-\(d\) interaction strength is given by 
$
J_{sd} \approx \frac{V_t^2}{E_m}
$
where \(E_m\) represents the energy difference between the quantum dot (denoted as \(E_0\)) and the Fermi energy, specifically \(E_m = E_F - E_0\). 
\(V_t\) is the tunneling matrix element in the Anderson Hamiltonian.
Then, we have
\begin{eqnarray}
J_{23}
&\approx & \frac{a_0^2R_y }{16\pi m^*} \left(\frac{\Gamma}{E_m}\right)^2 k_F^2 F'_3(2k_FW),
\end{eqnarray}
where $a_0=0.00529$ nm, $R_y=13.606$ eV, and $m^*=0.2$.
$\Gamma$ represents the tunneling coupling between 
the qubit and channel.
Figure~\ref{fig_concept}(c) shows an example of the numerical estimation of the interaction.

\section*{Acknowledgment}
T. Tanamoto is grateful to S. Takagi, T. Mori, and H. Fuketa for fruitful discussions.
We are grateful to R. Honda for his technical support with the Silvaco TCAD Simulator Atlas. 
We are also grateful to Y. Yamamoto for their support with SmartSpice. 
We also thank Grammaly and Editage for English language editing.

\end{document}